\documentclass[aps,prl,reprint, groupedaddress,showpacs]{revtex4-1}

\usepackage{amsmath, amssymb, amsfonts}

\usepackage{epsfig}
\usepackage{epstopdf}

\usepackage{bm}
\usepackage{graphicx}
\usepackage{xcolor}
\usepackage{soul}

\renewcommand{\d}{\dag}
\newcommand{\h}{\hat}

\newcommand{\be}{\begin{equation}}
\newcommand{\ee}{\end{equation}}
\newcommand{\bi}{\begin{itemize}}
\newcommand{\ei}{\end{itemize}}
\newcommand{\ben}{\begin{enumerate}}
\newcommand{\een}{\end{enumerate}}

\newcommand{\beq}{\begin{equation}}
\newcommand{\eeq}{\end{equation}}
\newcommand{\bea}{\begin{eqnarray}}
\newcommand{\eea}{\end{eqnarray}}

\newcommand{\<}{\langle}
\renewcommand{\>}{\rangle}

\newcommand{\commentout}[1]{{}}
\newcommand{\half}{\hbox{$1\over2$}}
\newcommand{\eq}[1]{Eq.~\eqref{#1}}

\begin{document}

\title{Local and spatially extended sub-Poisson atom number fluctuations in optical lattices}
\author{C. Gross} \affiliation{Kirchhoff-Institut f\"ur Physik, Universit\"at Heidelberg,
Im Neuenheimer Feld 227, 69120 Heidelberg, Germany}
\author{A. D. Martin}
\affiliation{School of Mathematics, University of Southampton,
Southampton, SO17 1BJ, UK}
\author{J. Est\`eve} \altaffiliation{Present address: Laboratoire Kastler Brossel, CNRS, UPMC, Ecole Normale Sup\'erieure, 24 rue Lhomond, 75231 Paris, France.}
\author{M. K. Oberthaler}
\affiliation{Kirchhoff-Institut f\"ur Physik, Universit\"at Heidelberg,
Im Neuenheimer Feld 227, 69120 Heidelberg, Germany}
\author{J. Ruostekoski}
\affiliation{School of Mathematics, University of Southampton,
Southampton, SO17 1BJ, UK}

\begin{abstract}
We demonstrate that ultracold interacting bosonic atoms in an optical lattice with large on-site population show sub-Poissonian on-site and inter-site atom number fluctuations.  The experimental observations agree with numerical predictions of the truncated Wigner approximation.
The correlations persist in the presence of multi-mode atom dynamics and even over large spatially extended samples involving several sites.
\end{abstract}
\pacs{03.75.Lm,03.75.Gg,03.65.Ta,42.50.Lc}

\date{\today}
\maketitle

The preparation of ultracold ensembles with well-defined numbers of atoms is advantageous for high-precision measurements.
In quantum metrology~\cite{Bouyer:1997aa} the reduced uncertainty in the relative particle number between two ensembles is a prerequisite for spin-squeezing employed in quantum interferometric sensors \cite{SpinSqTheo, SqueezedInterferometry}. Interferometers using ultracold atoms also benefit from a well-defined total atom number, e.g., due to constant interaction-induced level shifts.

The basic phenomenon that repulsive interactions can lead to reduced atom number fluctuations has been observed in shallow traps \cite{Chuu:2005aa} and optical lattices~\cite{Esteve_Nature_2008, Sherson:2010aa,Bakr:2010aa, FluctIndirect,Gerbier:2006aa}. Lattice systems have been used for preparation of spin-squeezed states for quantum-enhanced interferometry~\cite{Esteve_Nature_2008}.
Here we report on the suppression of {\em absolute} (on-site) as well as {\em relative} atom number fluctuations below the limit for independent classical particles in a few-site optical lattice. We work in the large site-occupation regime that is different from the low-filling limit where, e.g., Mott-insulator physics can be studied. By site-resolved detection we observe reduced fluctuations after adiabatically changing the lattice parameters. This method provides a much lower limit for the absolute number fluctuations than recently developed dissipative approaches~\cite{Itah:2010aa,Whitlock:2010aa}.
Our numerical simulations show that the observed fluctuations cannot be explained by simple models where only one mode per lattice site is considered. A multi-mode treatment is necessary and our numerical solutions indicate the presence of quantum correlations over spatial regions of several sites which reveal themselves as spin squeezing between the considered regions.

\begin{figure}[htbp]
\centering
\includegraphics[width=0.95\columnwidth]{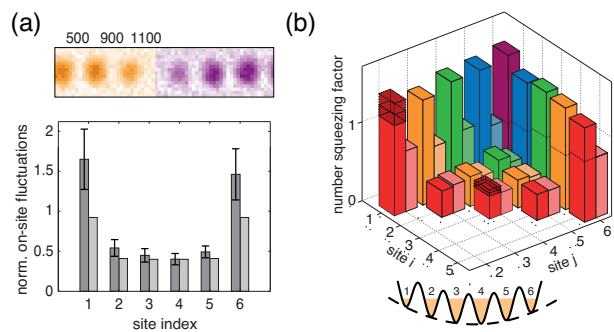}
\caption{ Spatially-resolved atom number fluctuations in an approximately symmetric six-site optical lattice after slow ramp up of the lattice potential. (a) Single-shot absorption image of the atomic cloud with typical mean atom numbers indicated. The experimentally observed normalized on-site fluctuations $\zeta^{2}$ (dark) are compared to TWA numerics (light). (b) Experimental (dark colors) and numerical (light colors) results for the relative atom number fluctuations $\xi^{2}_{N}$ between all site-pairs that increase with pair separation (different colors). Representative one standard deviation statistical error bars (transparent boxes in b) show increased experimental uncertainty for the outer wells. For this summarizing figure we average all measurements taken for final lattice heights larger than $900\,$Hz. } \label{Fig1}
\end{figure}
%
We load a Bose-Einstein condensate (BEC) of about $\langle N \rangle =5000$ $^{87}$Rb atoms into a 1D optical lattice resulting in  significant population of six sites (not analyzing two more sites with a population of less than $5\%$ of $N$).
Squeezed number fluctuations are observed after  slowly and linearly ramping up the lattice potential $V_l(x,t)=V(t) \sin^2(\pi x/d)$  (the ramp speed is ${\rm d} V/{\rm d}  t =4\,{\rm Hz/ms}$) from $V_{0} \simeq 430\,$Hz to $V \simeq 1500\,$Hz, with the lattice spacing $d\simeq 5.7\,\mu$m.
Some experiments were done with final lattice heights below $V \simeq 430\,$Hz in which case the ramp started from $V_{0} \simeq 250\,$Hz.
Additionally the atoms are held in a cigar-shaped trapping potential $V_{3D}(x,y,z)=m[\omega^2x^2+\omega_\perp^2(y^2+z^2)]/2$ with trap frequencies $\omega_\perp \simeq 2\pi\times 427\,$Hz and $\omega \simeq 2\pi\times21\,$Hz.
By single-site-resolved imaging we detect the local atom number $n_{i}$ and its fluctuations $\Delta n_{i}^2$ in each site $i$.

In Fig.~\ref{Fig1}a we plot the observed on-site fluctuations normalized to the fluctuations expected for a multinomial distribution. In the central sites we find  suppression of these fluctuations ($\zeta^2<1$,  defined below) due to repulsive interactions.
The increase of the variance towards the outer wells is qualitatively explained by thermal phonons being dominantly localized in those sites.
Figure~\ref{Fig1}b shows the spatially extended character of the number correlations across the lattice by considering relative atom number fluctuations $\xi_{\rm N}^{2}$ (defined below) between all well-pairs. We measure strongly suppressed relative fluctuations for all combinations of the central four wells. We also show numerical results of non-equilibrium dynamics based on
the truncated Wigner approximation (TWA) \cite{DRU93, Isella_PRA_2006}
(as detailed below) providing good agreement with the experiment for the central well-pairs, but underestimating the outer well contributions.
This deviation might be explained by experimental noise and the 1D nature of the numerical model which cannot exactly reproduce the physics in the experimentally realized crossover regime between 1D and 3D. It is important to note that simpler models, which neglect the on-site multi-mode structure, do not reproduce the experimental data.

\begin{figure}[tbp]
\centering
\includegraphics[width=0.95\columnwidth]{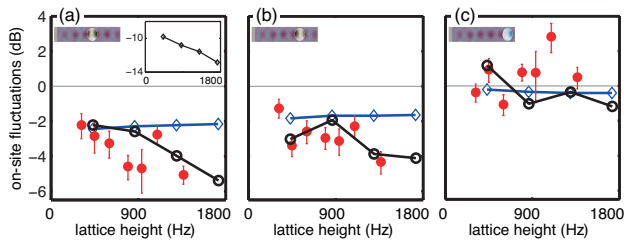}
\caption{Evolution of the on-site fluctuations during the lattice ramp. Experiment (red), eight-site discrete model  at $T\simeq16\,$nK (blue line), advanced model (TWA) at  $T\simeq 5.5\,$nK (black circles) and $T=0$ (inset). (a)-(c) show $\zeta^{2}$ in sites 4-6, respectively. Temperatures are chosen for the best agreement in all experimental observables shown in this manuscript. }  \label{Onsite}
\end{figure}

The evolution of the on-site fluctuation with lattice height is shown in Fig.~\ref{Onsite} for three representative wells. Between different experimental runs the integral atom number $N$ varies  by $\Delta N \simeq 1.6 \sqrt{\langle N \rangle}$. This technical noise is compensated by normalizing all data to the measured $N$ in each shot.
Due to possible long term drifts of the lattice position we calculate $p_{i} = \langle n_{i}/N\rangle$, the multinomial probability for well $i$, for each dataset consisting of $25$-$40$ experimental realizations. The on-site fluctuations $\zeta^{2}=\langle  n_{i,{\rm(corr)}}^{2}\rangle/p_{i}(1-p_{i})N$ are then obtained from the corrected site population $n_{i,{\rm(corr)}}=n_{i} - p_{i}N$ and normalized to the variance for a multinomial distribution. The results are averaged over a few datasets such that approximately $270$ experimental runs contribute to each data point and the uncertainty is estimated as the standard deviation of the different datasets.
Photon shot-noise in the detection process adds to the measured fluctuations which we subtract from the experimental data~\cite{Esteve_Nature_2008}.
With increasing lattice depth the central wells  show stronger suppression of $\zeta^{2}$, but
the fluctuations in the outer wells do not decrease significantly with lattice height which can be explained by the localization of the lowest phonon modes mainly on the edges.

In future experiments different lattice sites might be used as independent mesoscopic gases with highly suppressed fluctuations of their total atom number. Therefore we analyze the local fluctuations without correcting for fluctuations of $N$, the integral atom number,  and reference the fluctuations to the shot-noise limit of individual samples. We measure a four-fold suppression of the local fluctuations $\Delta n_{i}^{2}/\langle n_{i} \rangle$ in the central wells as compared to the normalized variance of the integral atom number $\Delta N^{2}/\langle N \rangle$. The fluctuations are even sub-Poissonian ($\Delta n_{i}^{2}/\langle n_{i} \rangle \simeq -2.2\,{\rm dB}$), showing that this experimental method can be used to prepare individually addressable samples with well-defined atom numbers~\cite{Chuu:2005aa, Whitlock:2010aa, Itah:2010aa}.

\begin{figure}[tbp]
\centering
\includegraphics[width=0.95\columnwidth]{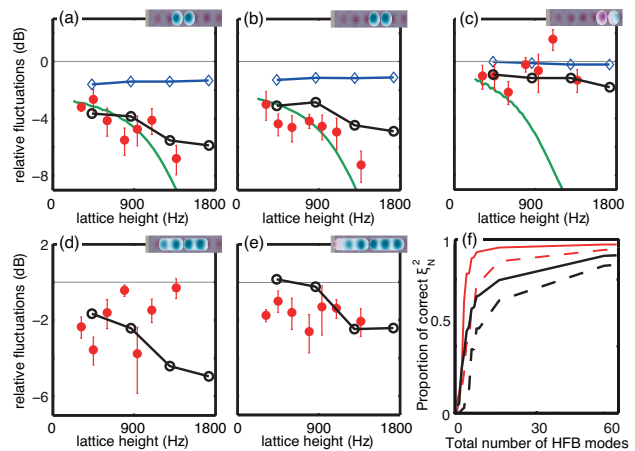}
\caption{Normalized relative atom number fluctuations $\xi^{2}_{N}$ versus lattice height for experiment (red circles), TWA numerics at $T\simeq5.5\,$nK (black open circles), the local two-mode model at $T\simeq 10\,$nK (green curve), eight-site discrete model at $T\simeq16\,$nK (blue line). (a)-(c) correspond to the well-pairs (34), (45) and (56). At bottom row $\xi^{2}_{N}$ between larger subsystems of adjacent sites binned together; (d) sites (\{23\}\{45\}); (e) sites (\{123\}\{456\}); (f) the cumulative contribution of the number of continuum HFB phonon modes to the correct value of $\xi_N^2$ for the pairs (34) (dashed line) and (\{123\}\{456\}) (solid line) at $T\simeq 5.5\,$nK (red) and $T=0$ (black).} \label{FiniteT}
\end{figure}

To probe spatially extended correlations we analyze the \emph{relative} atom number fluctuations between two lattice sites and between larger subregions of the lattice when the atom numbers of the individual sites are added together. In Fig.~\ref{FiniteT}a-c we show the evolution of relative atom number fluctuations between adjacent sites, normalized to the binomial variance: for the sites $i,j$, $\xi^{2}_{\rm N}={\rm var}(n_{i}-n_{j}) /4 p_{ij}(1-p_{ij})\langle n_{i}+n_{j}\rangle$, where $p_{ij} = \langle n_{i}/(n_{i}+n_{j})\rangle$.
The experimental results are limited by finite temperature effects as in the case of on-site fluctuations.
Remarkably, $\xi^{2}_{\rm N}$ are also squeezed between spatial regions incorporating more sites (Fig.~\ref{FiniteT}d,e), indicating how squeezing can be obtained even over extended lattice regions.

To identify the physical processes that are important in the observations we describe the system by different models with increasing complexity.
The simplest conceivable model is a local two-mode discrete model $\h H=E_{C} (\h n_{i}-\h n_{j})^{2}/8 - E_{J} (\h a_i^\dagger \h a_j +\h a_j^\dagger \h a_i) /N+ \Delta ( \h n_{i}- \h n_{j})/2$, as used in the analysis of spin squeezing by Est\`eve et al.~\cite{Esteve_Nature_2008}.
Our system is far from the tight-binding regime and the parameters are calculated from the ground state wavefunctions obtained from the 3D Gross-Pitaevskii equation (GPE) following the recipe in Ref.~\cite{ZAP98}. For the shallow lattice heights the tunneling timescale is in the order of a few tens of ms while its rate of change due to the lattice ramp is more than a factor of $10$ smaller. Figure~\ref{FiniteT}a-c reveals that this simple model can be fitted to the experimental observations for the central wells by choosing the initial temperature $T \simeq 10\,{\rm nK}$. For these wells the evolution is adiabatic within this model.
However, assuming constant $T$ throughout the lattice the fluctuations in the outer wells are not explained (Fig.~\ref{FiniteT}a-c).

A more complex model yielding the on-site fluctuations is the discrete multi-site model $\h H=-\sum_{<i,j>} J_{ij}(\h a_i^\dagger \h a_j +\h a_j^\dagger \h a_i)+\sum_j [\nu_j \h n_j  + \half U_{j}\h n_j (\h n_j-1)]$ for which we calculate
the Hartree-Fock-Bogoliubov (HFB) modes~\cite{Hutchinson_PRL_1998}.
The results are obtained by including eight sites in which case the outermost sites are almost empty -- adding more sites does not change the results.
The parameters $U_j,\nu_j$ are deduced analogously to the local two mode analysis.
Evaluation of the hopping amplitudes $J_{ij}$ is sensitive to numerical accuracy, and thus we take $J_{ij}$ for the central sites from the calculated $E_J$ and adjust the remaining $J_{ij}$'s  such that the ground state populations match with the experiment. Figure~\ref{Onsite} shows the on-site fluctuations providing the best agreement with the experiment when taking the initial temperature as a free parameter and assuming adiabatic lattice ramping.
The results qualitatively reproduce the level of experimentally observed on-site fluctuations across the lattice, but the corresponding values for $\xi_N^2$ are slightly higher than the experiment in Fig. 3 and the trend of decreasing fluctuations with lattice height is not predicted by this model.

The failure of the discrete models to reproduce all experimental observations indicates the importance of multimode effects. To reveal these contributions we describe the system by a continuum multi-band 1D Hamiltonian $ H=\int dx [\h\psi^\dagger(T +V-\mu)\h\psi+ g_{1D}
\h\psi^\dagger\h\psi^\dagger\h\psi\h\psi/2]$, for which we first calculate the equilibrium state ($T$, $V$, and $g_{1D}$ denote the kinetic energy, potential, and nonlinearity).
We find the Bogoliubov theory inaccurate, in which case the back-action of the excited-state population on the ground state is ignored, and we consequently use the gapless HFB formalism  where the ground-state
and the excited-state correlations are solved self-consistently~\cite{Hutchinson_PRL_1998}.
Specifically, for Bogoliubov and HFB modes we define the number operator in the site $i$ as
$
\hat{n}_i = \int_i dx\, \h\psi^\dagger (x) \h\psi(x)
$
where the integration is over the $i$th site. The
site population is then $\<\hat{n}_i\> = N_0 \int_i dx\, \psi_0(x)^2 + \int_i dx\,
\<\hat{\delta}^\dagger(x)\hat{\delta}(x)\>$, where $\h\psi(x)=\psi_0(x)\h\alpha_0+ \hat{\delta}(x)$  with
$\<\h\alpha_0^\dagger\h\alpha_0\>=N_0$ and the excitations $\hat{\delta}(x)= \sum_{j>0} \big[
u_j(x)\h\alpha_j-v^*_j(x)\h\alpha^\d_j \big]$. The atom number fluctuations in the $i$th lattice site $\delta n_{i}^{2}=\langle \hat{n}_{i}^2\rangle-\langle \hat{n}_{i}\rangle^2$ are evaluated analogously
\begin{align}
\delta n_{i}^{2}&\simeq N_0\sum_{j}(2N_{j}+1)\left|\int_{i^{\rm th} {\rm well}} dx\psi_{0}\left[u_{j}(x)-v_{j}(x)\right]\right|^{2}\nonumber \\
&+\int_{i^{\rm th} {\rm well}} dx dx'\left[\langle \hat{\delta}^{\dagger}(x)\hat{\delta}^{\dagger}(x')\rangle\langle\hat{\delta}(x)
\hat{\delta}(x')\rangle
\right.\nonumber \\ &\left.\quad\quad\quad\quad\quad\quad+\langle \hat{\delta}^{\dagger}(x)\hat{\delta}(x')\rangle\langle\hat{\delta}(x)\hat{\delta}^{\dagger}(x')
\rangle \right]\,.\label{Eq_dn}
\end{align}
Compared to HFB, the Bogoliubov method overestimates both the linear and quadratic terms of $\delta n_i^2$ [the first and the second term in Eq.~(1)]. For the shallowest lattice the difference in $\delta n_{i}^{2}$ is close to 10\% (over 100\%) at at $T\simeq5.5\,$nK ($8\,$nK).
The quadratic contribution at $T\simeq5.5\,$nK is more than two times larger than the linear one indicating the importance of the phonon-phonon interactions on the number fluctuations even within individual sites. The large number of modes required to describe the fluctuations emphasizes multi-mode quantum effects in individual sites as opposed to the single mode per site in the tight-binding model. This is visualized in Fig.~\ref{FiniteT}f where we show the cumulative contribution of the number of continuum HFB modes to atom number fluctuations, derived from expressions analogous to \eq{Eq_dn}.

In order to model the non-equilibrium dynamics of the experiment in the continuum multi-band theory
we employ the 1D TWA. In the TWA implementation we otherwise follow Ref.~\cite{Isella_PRA_2006}, except that we develop a projection method capable of analyzing the multi-mode effects in each site and use the multi-band HFB modes in TWA to provide a more accurate description for the initial state
quantum noise.
Although the experiment is not strictly 1D, the axial modes provide the main contribution to the observed number fluctuations.
Previous theoretical studies on atom number fluctuation dynamics have either concentrated on double-well systems \cite{Ferris_2010} or multi-well systems effectively in the tight-binding limit \cite{Isella_PRA_2006,SCH06,STR07}, so that significant intra-site multi-mode effects were absent.
In TWA the quantum field operator $\hat{\psi}$ is replaced by an ensemble of stochastic fields $\psi_{W}$ satisfying the  GPE \cite{Isella_PRA_2006}.
The Wigner representation returns symmetrically-ordered expectation values that we transform to normally-ordered ones by constructing an eigenmode basis in each individual site and projecting the field operator to the several lowest modes at different lattice heights
$s(t)$ during the turning-up of the lattice. We define the amplitude of the
$k$th mode of the site $i$ as
\beq
\label{projection}a_{i,k}(t)=\int_{i^{\rm th} {\rm well}}dx\,
[\varphi_{i,k}(x,t)]^*\psi_W(x,t)\,,
\eeq
where $\psi_W$ is the stochastic field
and $\varphi_{i,k}$ is the $k$th localized mode function
of the well $i$. Then the site population reads
\beq
\<\h n_i\>= \sum_k \< \h a_{i,k}^\dagger \h a_{i,k}\>=\sum_k \big[ \< a_{i,k}^*  a_{i,k}\>_W-1/2 \big]
\eeq
where $\<\cdots\>$ denotes the normally ordered expectation
value of the quantum operators, and $\<\cdots\>_W$ the symmetrically-ordered expectation values obtained from the TWA simulations. Fluctuations are calculated using analogous transformations
and the phase in each site is defined as a spatial average of the (slowly varying) multi-mode field $\phi_i\simeq{\rm arg}\int_{i}dx\,\sum_k a_{i,k}\varphi_{i,k}(x)$.
We first let the initial state in TWA to evolve in a stationary lattice with $V_{0} \simeq 430\,$Hz,
in order to damp out excitations due to the approximate nature of the initial HFB solution.  We then turn the lattice up calculating the number squeezing during the ramping and average over remaining small oscillations in the observables. The black lines in Figs.~\ref{Onsite}-\ref{FiniteT} show the TWA results which reproduce the experimentally observed behavior for $T\simeq5.5 (0.5)\,$nK.
The TWA results are very sensitive to $T$, however, its absolute value can not be expected to match with the temperature in the experiment due to the different dimensionality.

Building on the successful comparison between theory and experiment we discuss the implications for spin-squeezing in this system.
A sufficient condition for the initial state of quantum-enhanced precision measurement and many-body entanglement is that the spin-squeezing parameter between two subsystems $\xi_S^2= N (\Delta \h S_z)^2/(\<\h S_x\>^2+\<\h S_y\>^2)<1$~\cite{Sorensen:2001aa}, where the pseudo spin operator $\h S_z$ -- proportional to the population difference in the two subsystems -- represents the squeezed quadrature and the mean spin is aligned along the $xy$ plane. Next to number squeezing $\xi_N^{2}$, which measures $\Delta \h S_{z}^{2}$, the coherence $(\<\h S_x\>^2+\<\h S_y\>^2)$ is an important parameter here.
For adjacent sites $\xi_S^{2}\simeq \xi_N^{2}/\<\cos \phi\>^{2}$ where the coherence is experimentally observable by local interference measurements~\cite{Esteve_Nature_2008} after ballistic expansion~\cite{infcomment}.
The measurements yielded $\xi_S^{2}\simeq -3.8\,$dB for the central sites. The experimentally measured phase coherence is notably lower than $\<\cos \phi\>^{2}$ obtained from the argument of the TWA wave function (that incorporates contributions from the entire momentum distribution), indicating that spin-squeezing may even be stronger than the experimentally inferred value. We numerically simulated the effect of the 3D ballistic expansion, but the expansion itself cannot explain the lower experimental values.
At $T\simeq5.5\,$nK and $V\simeq1750\,$Hz we calculate $\xi_S^{2}\simeq-5.5\,$dB for the site-pairs (3-4) while the value at $T=0$ is $\xi_S^{2}\simeq -12.5\,$dB.
Numerically we also found notable spin-squeezing between non-nearest-neighbor site-pairs and between the extended lattice regions examined above.

In conclusion we prepared sub-shot noise ultracold atomic samples with large atom numbers in a few-site optical lattice. The observations are explained by 1D TWA simulations. By comparison with simple discrete models we find that only the advanced TWA model explains all observables consistently revealing the importance of multi-mode effects and also providing an additional confirmation of the observations.
The temperatures used to match the different models to the experiment vary significantly, indicating that temperature measurements based on fluctuations in this complex system are model dependent.
The numerical studies show that taking into account phonon-phonon interactions is crucial and furthermore predict that spin-squeezing can be generated over extended lattice regions.

\begin{acknowledgements}
We gratefully acknowledge fruitful discussions  with S.\ Giovanazzi and
support from the Heidelberg Center of Quantum Dynamics, DFG Forschergruppe 760, the German-Israeli Foundation, the ExtreMe Matter Institute, the European Commission FET open scheme project MIDAS, and EPSRC.
\end{acknowledgements}

\end{document}